\documentstyle[12pt,psfig]{article}
\begin{document}
\begin{center}
{\Large\bf Cosmic acceleration in non-canonical scalar field model: an interacting scenario}
\\[15mm]
Sudipta Das \footnote{E-mail:sudipta.das@visva-bharati.ac.in}~~and 
Abdulla Al Mamon \footnote{E-mail : abdullaalmamon.rs@visva-bharati.ac.in}\\
{\em Department of Physics, Visva-Bharati,\\
Santiniketan- 731235, ~India.}\\
[15mm]
\end{center}

\vspace{1cm}

\pagestyle{myheadings}
\newcommand{\be}{\begin{equation}}
\newcommand{\ee}{\end{equation}}
\newcommand{\bea}{\begin{eqnarray}}
\newcommand{\eea}{\end{eqnarray}}
\newcommand{\bc}{\begin{center}}
\newcommand{\ec}{\end{center}}

\begin{abstract}
In this paper we have studied the dynamics of accelerating scenario within the framework of scalar field models possessing a non-canonical kinetic term. In this toy model, the scalar field is allowed to interact with the dark matter component through a source term. We have assumed a specific form for the coupling term and then have studied the dynamics of the scalar field having a constant equation of state parameter. We have also carried out the dynamical system study of such interacting non-canonical scalar field models for power law potentials for some physically relevant specific values of the model parameters. It has been found that the only for two particular stable fixed points of the system, an accelerating solution is possible and the universe will settle down to a $\Lambda$CDM universe in future and thus there is no future singularity in this model. 
\end{abstract}
\section{Introduction}
Recent cosmological observations \cite{1,1a} strongly suggest that the present cosmic acceleration is driven by the mysterious dark energy which has large negative pressure i.e., long range anti-gravity properties. Several dark energy models have been constructed to explain this late time acceleration of the universe using a scalar field, namely, quintessence \cite{quin2,quin1,quin3, quin4, quin5},  having different forms of equation of state $p = w \rho$ for the dark energy \cite{gas,gas1}. A number of  other models also exist in the literature in which the action is modified in various ways. To name a few, there are non-minimally coupled scalar field models \cite{nonminimal,nonminimal3,nonminimal4,nonminimal5,nonminimal2,nonminimal1}, $f(R)$-gravity models \cite{fR2,fR,fR3,fR1}, non-canonical scalar field models \cite{noncan1,noncan11,noncan10b,noncan11a,noncan9,noncan5,noncan10,noncan8,noncan9a,noncan3,noncan2,noncan6,noncan7,
noncan10a,noncan4}, $k$-essence models \cite{kess1,kess2,kess} and many more. All these proposals offer quite  satisfactory description of the dark energy properties in some sense or the other although each of them have its own demerit. As none of these models can be considered as superior to others,  the search is on for a suitable dark energy candidate.\\ 
Most of the dark energy models consider the field to be non-interacting. As nothing specific is known about the nature of dark energy, an interacting scenario may be useful and may provide a more general scenario. Recently, interacting dark energy models have gained interest and there are several works in which investigations are carried out considering an interaction between the different components of the universe \cite{int,int1,int2,int3,int4,int5}. It has been found that an interacting scenario can provide solution to a number of cosmological problems in both canonical scalar field models \cite{sd0} as well as non-minimally coupled scalar field models such as Brans-Dicke scalar field \cite{sd1, sd2}. Motivated by these facts, in this paper we try to build up a viable dark energy model using a non-canonical scalar field in an interacting scenario.\\ 
\par In general, the Lagrangian for a scalar field model can be represented as \cite{Melch}
\be\label{Lgen}
L = f(\phi) F(X) - V(\phi)
\ee                             ̇
where $X = \frac{1}{2} {\dot{\phi}}^2$ for a spatially homogeneous scalar field. Equation (\ref{Lgen}) infact includes all the popular single scalar field models. It describes $k$-essence when $V(\phi) = 0$ and
quintessence when $f(\phi) =\rm{constant}$ and $F(X) = X$. In this paper, we consider another class of models called {\it General Non-Canonical Scalar Field Model} with its Lagrangian given by $L = F(X) - V(\phi)$.  This type of non-canonical scalar field models were proposed by Fang et al. \cite{Fang},  where they could obtain cosmological solutions for different forms of $F(X)$. Recently Unnikrishnan et al. \cite{unni} have proposed an inflationary model for the universe using a non-canonical scalar field.\\
\par Following \cite{Fang} and \cite{unni}, we consider a non-canonical scalar field cosmological model in the framework of a homogeneous, isotropic and spatially flat FRW space-time. We also consider that the non-canonical scalar field is interacting and we introduce a source term through which the scalar field interacts with the matter field. Under this scenario, we derive the field equations and try to obtain exact solutions for various cosmological parameters.  
A brief description of the paper is as follows: In section 2, we have derived the field equations and the conservation equations for the non-canonical scalar field Lagrangian. We have shown that with a constant equation of state parameter for this non-canonical scalar field, this interacting model can provide an accelerated expansion phase of the universe preceded by a decelerating phase. We could obtain cosmological solutions for various parameters of the model. In section 3, we perform the dynamical system study for this interacting non-canonical scalar field model of the universe with simple power law potentials. The last section contains some concluding remarks.
\section{Field equations and their solutions}
\begin{figure}[!h]
\begin{center}
\centerline{\psfig{figure=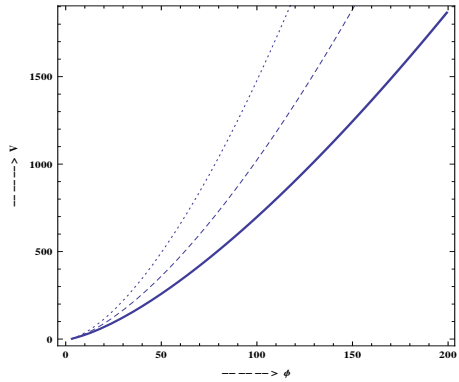,height=50mm,width=60mm}}
\caption{\normalsize{\em Plot of $V$ vs. $\phi$ for $\epsilon=1.1$ (thick curve), $\epsilon=1.2$ (dashed curve), $\epsilon=1.3$ (dotted curve), $\omega = -0.9$, $V_{0} = 2$, $B=0.3$ and $\phi_{0} = 3$.}}
\label{figv1}
\end{center}
\end{figure}

Let us consider the action
\be\label{action}
S = \int\sqrt{-g} dx^{4}\left[\frac{R}{2} + {\cal L}(\phi,X)\right] + S_{m}
\ee
(We have chosen the unit where $8{\pi}G=c=1$.)\\
where $R$ is the Ricci scalar, $S_{m}$ represents the action of the background matter, the Lagrangian density ${\cal L}(\phi,X)$ is an arbitrary function of the scalar field $\phi$, which is a function of time only and its kinetic term $X$ given by $X = \frac{1}{2}{\partial_{\mu}}\phi{\partial^{\mu}}\phi$. 
\vspace{3mm}\\
Variation of the action (\ref{action}) with respect to the metric $g^{\mu\nu}$ gives the Einstein field equations as
\be
G_{\mu\nu} = \left[T^{\phi}_{\mu\nu} + T^{m}_{\mu\nu}\right]
\ee
where the energy-momentum tensor of the scalar field $\phi$ is given by
\be
T^{\phi}_{\mu\nu} = \frac{\partial {\cal L}}{\partial X}{{\partial_{\mu}}\phi {\partial_{\nu}}\phi} - g_{\mu\nu}{\cal L}.
\ee
$T^m_{\mu\nu}$ represents the energy-momentum tensor of the matter component which is modelled as an idealized perfect fluid, and is given by
\be
T^m_{\mu\nu} = (\rho_{m} + p_{m})u_{\mu}u_{\nu} + pg_{\mu\nu}
\ee 
where ${\rho}_{m}$ and $p_{m}$ are the energy density and pressure of the matter components of the universe respectively. The four-velocity of the fluid is denoted by $u_{\mu}$. Also, variation of the action (\ref{action}) with respect to the scalar field $\phi$ gives the equation of motion for $\phi$ field as
\be
{\ddot{\phi}}\left[\left(\frac{\partial {\cal L}}{\partial X}\right) + (2X)\left(\frac{\partial^2 {\cal L}}{\partial X^2}\right)\right] + \left[3H \left(\frac{\partial {\cal L}}{\partial X}\right) + {\dot{\phi}}\left(\frac{\partial^2 {\cal L}}{\partial X\partial \phi}\right)\right]{\dot{\phi}} - \left(\frac{\partial {\cal L}}{\partial \phi}\right) = 0 
\ee\\
Let us consider a homogeneous, isotropic and spatially flat FRW universe which is characterized by the line element
\be\label{metric}
ds^{2} = dt^{2} - a^{2}(t)[dr^{2} + r^{2}d{\theta}^{2} +r^{2}sin^{2}\theta d{\phi}^{2}]
\ee
where $a{\rm{(t)}}$ is the scale factor of the universe with cosmic time $t$. Here, we only consider the spatially flat FRW universe as indicated by the anisotropy of the CMBR measurement \cite{pde}. The Einstein field equations for the space-time given by equation (\ref{metric}) with matter in the form of pressureless perfect fluid (i.e., $p_{m} = 0$) takes the form, 
\be\label{eq1}
3H^{2} = {\rho}_{m} + {\rho}_{\phi},
\ee
\be\label{eq2}
2{\dot{H}} + 3H^{2} = -p_{\phi}
\ee
\be\label{eq3}
{\dot{\rho}}_{\phi} + 3H(\rho_{\phi} + p_{\phi}) = 0
\ee
\be\label{eq4}
{\dot{\rho}}_{m} + 3H{\rho}_{m} = 0
\ee
Here $H = \frac{\dot{a}}{a}$ is the Hubble parameter and an overdot indicates differentiation with respect to the time-coordinate $t$. The expressions for the energy density $\rho_{\phi}$ and the pressure $p_{\phi}$ of the scalar field $\phi$ are given by
\be\label{rhopphi}
\rho_{\phi} = \left(\frac{\partial {\cal L}}{\partial X}\right)2X - {\cal L}, \hspace{5mm} p_{\phi} = {\cal L}
\ee
where $X = \frac{1}{2}\dot{\phi}^2$.\\
A number of functional forms of ${\cal L}(\phi, X)$ have been considered in the literature, see for instance Refs. \cite{noncan1,noncan11,noncan10b,noncan11a,noncan9,noncan5,noncan10,noncan8,noncan9a,noncan3,noncan2,noncan6,noncan7,
noncan10a,noncan4}. In this present work, we choose a Lagrangian density of the following form :
\be\label{L}
{\cal L}(\phi,X) = X^2 - V(\phi)
\ee
where $V(\phi)$ is the potential for the scalar field $\phi$. \\
Equation (\ref{L}) can be obtained from the general form of Lagrangian density 
\be\label{L1}
{\cal L}(\phi,X) = X \left(\frac{X}{{M_p}^4}\right)^{\alpha - 1} - V(\phi)
\ee
considered by several authors \cite{noncan6, unni, sanil} for $\alpha = 2$ and $M_p = 1/\sqrt{8\pi G} =1$. It is interesting to note that for $\alpha = 1$, equation (\ref{L1}) reduces to the usual Lagrangian density for a canonical scalar field model $\left[{\cal L}(\phi,X) = \frac{1}{2}\dot{\phi}^2 - V(\phi) \right]$.\\
From equation (\ref{rhopphi}), one can obtain the energy density and the pressure for the $\phi$-field corresponding to the above Lagrangian density as 
\be\label{rhophi}
\rho_{\phi} = 3X^2 + V(\phi), {\hspace{5mm}}p_{\phi} = X^2 - V(\phi)
\ee
with
\be\label{X}
X = \frac{1}{2}{\dot{\phi}}^{2}
\ee
Considering equations (\ref{L}), (\ref{rhophi}) and (\ref{X}), equations (\ref{eq1}) and (\ref{eq2}) take the form 
\be\label{eqx1}
3H^{2} = {\rho}_{m} + \frac{3}{4}{\dot{\phi}}^4 + V(\phi),
\ee
\be\label{eqx2}
2{\dot{H}} + 3H^{2} = -\frac{1}{4}{\dot{\phi}}^4 + V(\phi)
\ee
Also the conservation equations (\ref{eq3}) and (\ref{eq4}) take the form
\be\label{eqx3}
{\dot{\rho}}_{\phi} + 3H {\dot{\phi}}^4  = 0
\ee
\be\label{eqx4}
{\dot{\rho}}_{m} + 3H{\rho}_{m} = 0.
\ee
With the above set of equations (\ref{eqx1}) - (\ref{eqx4}), we want to build up an accelerating model for the universe in which the non-canonical scalar field plays the role of dark energy such that the universe smoothly transits from a decelerating to an accelerating phase. This is a must for the structure formation of the universe. This type of transition has been achieved in a number of canonical scalar field models \cite{sd0, sd},  modified gravity models or non-minimally coupled scalar field models \cite{sd1, sd2, sd3, am}. In this work, we try to achieve the same in non-canonical scalar field model of dark energy.\\
Usually the dark matter and scalar field components are considered to be non-interacting. But as nothing is known about the nature of the dark energy, an interaction between the two matter components will provide a more general scenario.  At this stage we consider that the scalar field and the dark matter components do not conserve separately but interact with each other through a source term (say $Q$). The sign of $Q$ will determine the direction of energy flow  between these components. The equations (\ref{eq3}) and (\ref{eq4}) under such a scenario generalize to
\be\label{rhoqp}
{\dot{\rho}}_{\phi} + 3H(1 + \omega_{\phi})\rho_{\phi} = Q
\ee
\be\label{rhoqm}
{\dot{\rho}}_{m} + 3H{\rho}_{m} = -Q
\ee
We choose a simple functional form of $Q$ as
\be\label{Q}
Q = \alpha H{\dot{\phi}}^{4} 
\ee
where $\alpha$ is an arbitrary constant and characterizes the strength of the coupling. There is no fundamental theory concerning the form of coupling term in the dark sector till now. In addition, we have lack of information regarding the nature of the dark sectors. So, if the dark sectors are allowed to interact among themselves, the coupling term $Q$ can be of any arbitrary form. It deserves mention that the ansatz in equation (\ref{Q}) is purely phenomenological and is motivated by mathematical simplicity.\\
Out of the four equations (equations (\ref{eq1}), (\ref{eq2}), (\ref{rhoqp}) and (\ref{rhoqm})), only three  are independent and the fourth one can be derived from the others using the Bianchi identity. So, we have three independent equations to solve for four unknown parameters $a$, $\rho_{m}$, $\phi$ and $V(\phi)$. Hence, one assumption can be made to match the number of unknowns with the number of independent equations.\\
It is well known observationally that at present $\omega_{\phi} \simeq -1$ \cite{vasey,davis} and the  condition for acceleration (${\ddot{a}} > 0$) gives the effective EoS parameter: $-1 <\omega_{\phi} < -\frac{1}{3}$. Motivated by these restrictions which shows that the range of allowed values of $\omega_{\phi}$ is very small, we consider that the EoS parameter $\omega_{\phi}$ is constant (say, $\omega$) and is given by the following simple relation
\be
\omega_{\phi} = \frac{p_{\phi}}{\rho_{\phi}} = \frac{X^2 - V}{3X^2 + V} = \omega ( \rm{a ~constant})
\ee
This makes the system of equations closed now.\\
The above equation leads to
\be\label{phidot}
X^2 = \frac{1+\omega}{1-3\omega} V {\hspace{5mm}}\Rightarrow  {\dot{\phi}}^{4}= 4\left(\frac{1+\omega}{1-3\omega}\right) V(\phi)
\ee
From equations (\ref{rhophi}) and (\ref{phidot}), we obtain the expressions for the energy density and pressure for this interacting model as 
\be\label{rhoxphi}
\rho_{\phi} = \frac{4}{(1 - 3\omega)}V(\phi),{\hspace{5mm}}p_{\phi} = \frac{4\omega}{(1 - 3\omega)}V(\phi) 
\ee
Using equation (\ref{rhoxphi}) and changing the argument from $t$ to $a$, equation (\ref{rhoqp}) can be re-written as
\be\label{V}
\frac{dV(a)}{da}{\dot{a}} + \epsilon \frac{\dot{a}}{a}V(a) = 0{\hspace{5mm}} \Rightarrow V(a) = V_{0}a^{-\epsilon}
\ee
where $V_{0}$ is a positive constant and $\epsilon = (3 - \alpha)(1 + \omega)$.\\
From equation (\ref{eq2}), (\ref{rhoxphi}) and (\ref{V}), by simple algebra one can obtain the form of the Hubble parameter for this model as
\be\label{h1}
H^2 = \gamma a^{-\epsilon} + Ba^{-3}
\ee
where $\gamma = \frac{4\omega V_{0}}{(3-\epsilon)(3\omega -1)}$ and $B$ is a positive integration constant.\\
Finally using equations (\ref{phidot}) and (\ref{V}), one can obtain the expression for evolution equation of the non-canonical scalar field $\phi$ as
\be\label{phixa}
\phi(a) = A \int {f(a)da} + \phi_{0}
\ee
where
\be
f(a) = {\frac{1}{a}}{\sqrt{\left( \frac{a^{3-\frac{\epsilon}{2}}}{1+\frac{\gamma}{B}a^{(3-\epsilon)}}\right)}}, 
\ee
$A = {\sqrt{\frac{2}{B}}}{\left[V_{0}\frac{1+\omega}{1-3\omega}\right]}^{\frac{1}{4}}$ and $\phi_{0}$ is the constant of integration. It is worth mentioning that for some specific choices of $\alpha$, $\omega$ and $B$, which will determine the value of $\epsilon$ and eventually will provide a specific functional form of $f(a)$, equation  (\ref{phixa}) can be integrated to obtain an analytical expression for $\phi (a)$. So it is found that in principle, a potential $V(a)$ leads to a number of possible solutions for the scalar field for different values of the model parameters. \\
The evolutionary trajectory for the potential $V(\phi)$ corresponding to the scalar field $\phi$ is shown in Figure \ref{figv1}. It has been found that the nature of the potential does not crucially depend on the values of $\epsilon$, $\omega$, $V_{0}$, $B$ etc. \\
From equation (\ref{eq1}), one can also obtain the expressions for the energy densities ${\rho}_{\phi}$ and $\rho_{m}$ as
\be\label{rhophiz}
\rho_{\phi}(z) = \frac{4}{(1 - 3\omega)}V = \frac{4V_{0}}{1-3\omega}(1+z)^{\epsilon}
\ee
\be\label{rhomz}
\rho_{m}(z) = 3H^2 - {\rho}_{\phi} = 
\left({3\gamma} - \frac{4V_{0}}{1-3\omega}\right)(1+z)^{\epsilon} + 3B(1+z)^{3}
\ee
where $1 + z = \frac{a_{0}}{a}$ and $a_{0} = 1$ is the present value of the scale factor. \\
Now, from equation (\ref{rhomz}) it is found that in order to ensure the positivity of the energy density for the matter field, the value of $\epsilon$ is constrained as, $\epsilon <3$ ( evident from the expression for $\gamma$). For plotting the figures, we have chosen the value of $\epsilon \sim 1.1$ or $1.2$ which is allowed value of $\epsilon$. \\
Also for the sake of completeness one can find out the expressions for the density parameters for the scalar field ($\Omega_{\phi}$) and the matter field ($\Omega_{m}$) as
\be\label{omegaphi}
\Omega_{\phi}(z) = \frac{4V_{0}}{3B(1-3\omega)}\left[\frac{(1 + z)^{\epsilon - 3}}{1 + \frac{\gamma}{B}(1 + z)^{\epsilon - 3}}\right],{\Omega_{m}(z) = 1 - \Omega_{\phi}(z)}
\ee
\begin{figure}[!h]
\begin{center}
\centerline{\psfig{figure=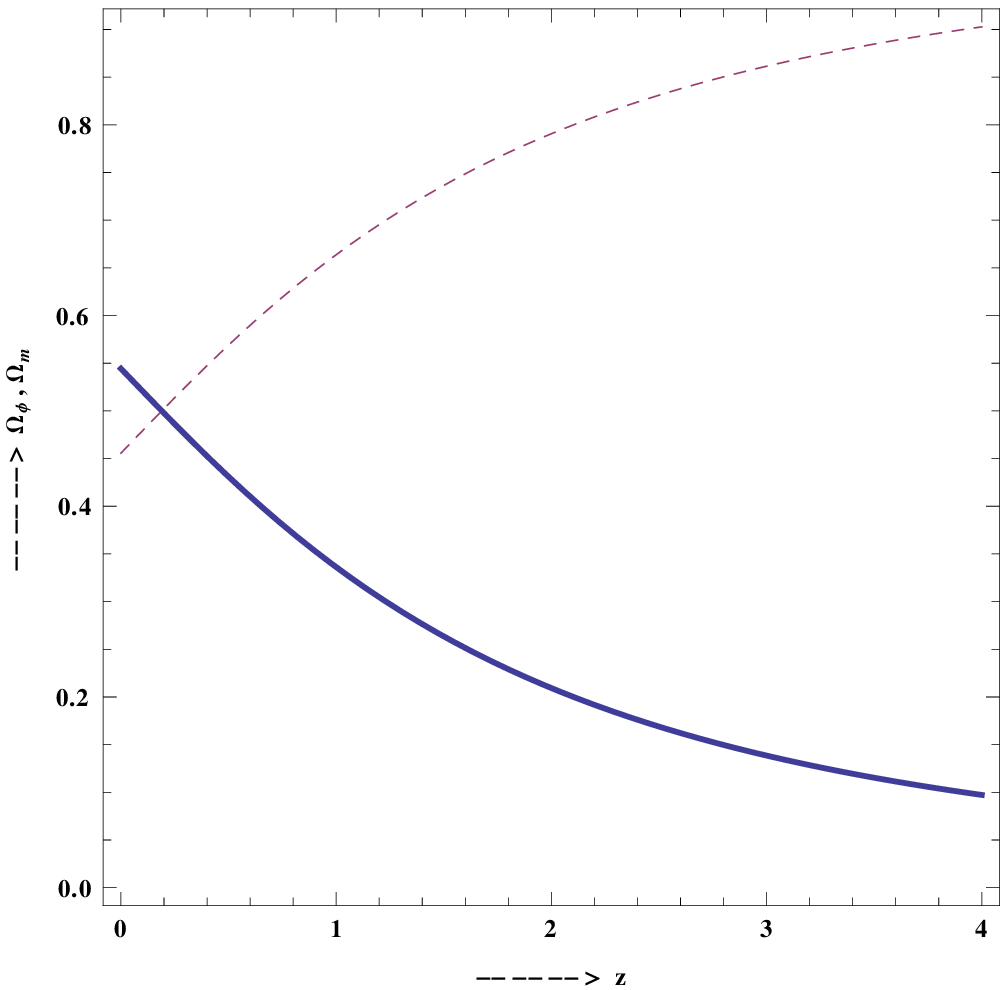,height=50mm,width=60mm}}
\centerline{\psfig{figure=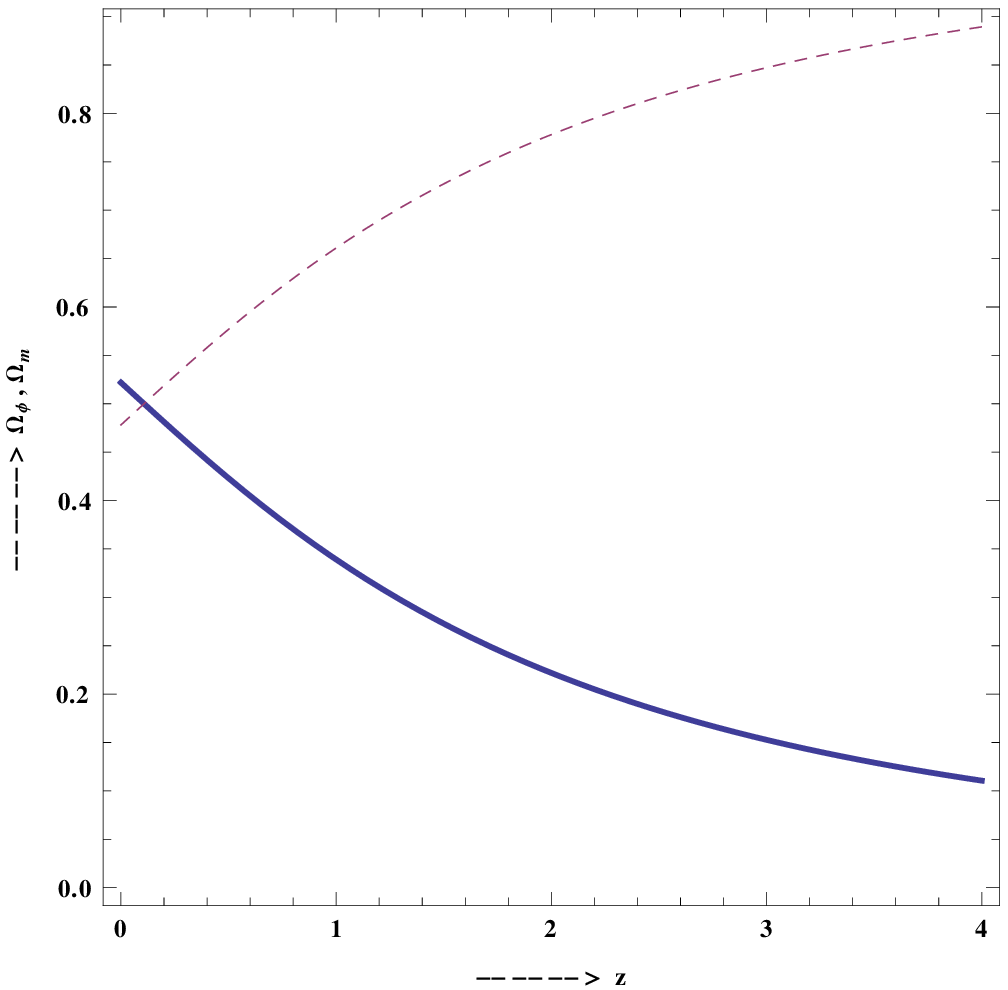,height=50mm,width=60mm}}
\caption{\normalsize{\em Plot of ${\Omega}_{m}$ {(dashed curve)} and  ${\Omega}_{\phi}$ {(solid curve)} as a function of $z$ for $\epsilon=1.1$ (upper panel) and $\epsilon=1.2$ (lower panel). We have chosen $\omega=-0.9$, $V_{0}=2$ and $B=0.3$. Here, $\gamma = \frac{4\omega V_{0}}{(3-\epsilon)(3\omega -1)}$.}}
\label{figomega1}
\end{center}
\end{figure}
Figure (\ref{figomega1}) shows the variation of density parameters for the two fields with redshift $z$. It shows that $\Omega_{\phi}$ remains sub-dominant in the early epoch and starts dominating in the recent past. It has been also found that the nature of the plot is insensitive to slight variation of the values of $\epsilon$, $B$ etc.\\
As mentioned earlier we are mainly interested in a model for the universe which smoothly transits from a decelerating to an accelerating phase. This can be best understood from the evolution of the deceleration parameter of the universe. The deceleration parameter $q$ for this model comes out as
\be
q(z) =-\frac{\ddot{a}}{aH^2} = -1 - \frac{\dot{H}}{H^2} = \frac{1}{2} + \frac{2\omega V_{0}}{B(1 - 3\omega)}\left[\frac{(1 + z)^{\epsilon - 3}}{1 + \frac{\gamma}{B}(1 + z)^{\epsilon - 3}}\right] 
\ee
\begin{figure}[!h]
\begin{center}
\centerline{\psfig{figure=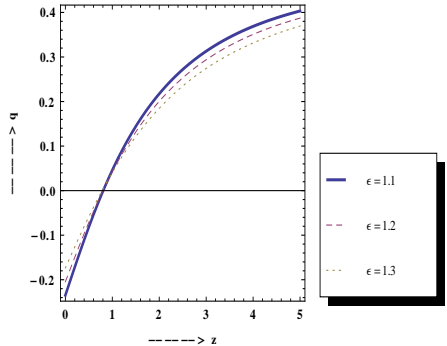,height=50mm,width=60mm}}
\caption{\normalsize{\em Plot of $q$ vs. $z$ for different values of $\epsilon$, $\omega = -0.9$, $V_{0} = 2$ and $B = 0.3$. Here, $\alpha = 3 - \frac{\epsilon}{1+\omega}$ and $\gamma = \frac{4\omega V_{0}}{(3-\epsilon)(3\omega -1)}$.}}
\label{figq1}
\end{center}
\end{figure}
It is clearly evident from figure (\ref{figq1}) that $q$ undergoes a smooth transition from a decelerated to an accelerated phase of the universe. The signature flip in $q$ takes place at around $z \approx 0.8$, which is in well agreement with the observations as suggested by \cite{Amendola}. The nature of the behaviour of $q$ against $z$ is also hardly affected by small change in the value of $\epsilon$ and other model parameters.\\
\par
\begin{figure}[!h]
\centerline{\psfig{figure=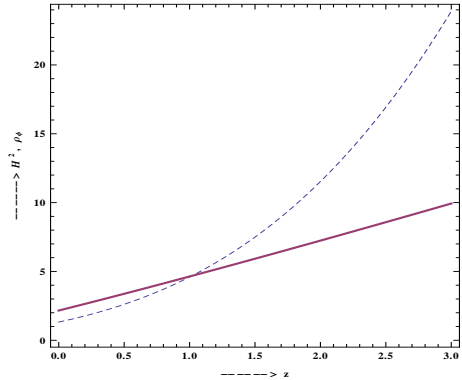,height=50mm,width=60mm}}
\caption{\normalsize{\em Plot of $\rho_{\phi}$ (solid line) and $H^2$ (dashed line) vs. $z$ for $\epsilon = 1.1$, $\omega = -0.9$, $V_{0} = 2$ and $B = 0.3$.}}
\label{figcompare}
\end{figure}

 {\bf Determination of sign of $Q$:}\\
For an accelerated universe, the EoS parameter $\omega$ must satisfy the inequality, $-1 < \omega < -\frac{1}{3}$ (except for tachyon models in which $\omega < -1$ is allowed). In our model, we have further constrained the value of $\epsilon = (3-\alpha)(1+\omega)$ as $\epsilon < 3$. As $\alpha$ and $\epsilon$  are related as $\alpha = 3 - \frac{\epsilon}{1+\omega}$, it is evident that $\alpha$ will be negative. As a consequence the parameter $Q=\alpha H {\dot{\phi}}^{4}$ also turns out to be negative for the present model. Equations (\ref{rhoqp}) and (\ref{rhoqm}) thus indicates that the energy gets transferred from the dark energy (DE) to the dark matter (DM) sector during the cosmic evolution. This is indeed counter intuitive as the dark energy component should dominate at later stages of cosmic evolution. A large number of DE models \cite{int,int1,int2,int3,int4,int5} have been proposed where energy flows from the DM sector to the DE sector such that the DE dominates over the DM at later times and drives the late-time acceleration of the universe. This provides a solution to the cosmological coincidence problem. In our model, however the direction of flow of energy is reverse. But recently Pavon and Wang \cite{pavon} have shown that as long as dark energy is amenable to a fluid description with a temperature not far from equilibrium, the overall energy transfer should be from DE to DM sector if the second law of thermodynamics and Le Chatelier-Braun principle are to be fulfilled. 	
\begin{figure}[!h]
\begin{center}
\centerline{\psfig{figure=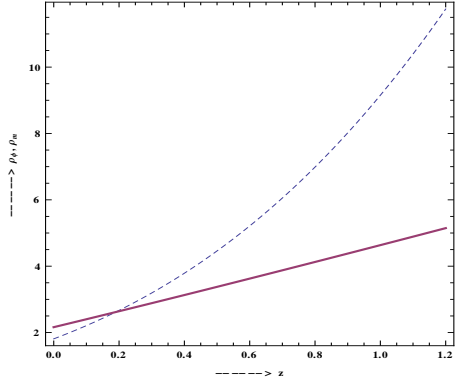,height=50mm,width=60mm}}
\caption{\normalsize{\em Plot of $\rho_{\phi}$ (solid line) and $\rho_m$ (dashed line) vs. $z$ for $\epsilon = 1.1$, $\omega = -0.9$, $V_{0} = 2$ and $B = 0.3$.}}
\label{figrho}
\end{center} 
\end{figure} 
\noindent
Also in the present model, we have the expressions for energy densities of the scalar field, $\rho_{\phi}$ (DE) and the normal matter field $\rho_m$ (DM) given by equations (\ref{rhophiz}) and (\ref{rhomz}) respectively. A plot of $\rho_{\phi}$ and $\rho_m$ vs. $z$ (Fig \ref{figrho}) shows that inspite of the fact that energy flows from the DE to the DM sector, the evolution dynamics of the universe for the present non-canonical model is such that the energy density of the DE sector dominates over the DM sector at late times. The reason behind this is that the rate of flow of energy is very less and thus energy density of the DM sector falls off more quickly than the DE sector. However it may happen that because of this reverse flow of energy, the DM sector may dominate over the DE sector in future and this accelerating phase of the universe may come to an end.\\ 
Also the plot of $(\Omega_m, \Omega_{\phi})$ vs. $z$ (Figure \ref{figomega1}) indicates that $\Omega_{\phi}$ increases as $z$ decreases. This is intriguing as we have obtained the direction of flow of energy from the DE to the DM sector. But figure \ref{figrho} shows that $\rho_{\phi}$ falls off with evolution. This unusual behaviour of $\Omega_{\phi}$ may arise if the square of the Hubble parameter $H^2$ falls off more rapidly than $\rho_{\phi}$ (since $\Omega_{\phi} = \frac{\rho_{\phi}}{3 H^2}$) which makes $\Omega_{\phi}$ increasing with the evolution. Figure \ref{figcompare} exhibits a similar behaviour where $\rho_{\phi}$ and $H^2$ are plotted as a function of $z$ for the same fixed values of the model parameters used in plotting figure \ref{figomega1}.     
\section {Dynamical system study}
In this section we rewrite the Einstein field equations for the interacting non-canonical scalar field model as a plane-autonomous system and study the stability of the critical points for the system. \\
For this purpose, we define three new variables :
\bc
$x = \frac{{\dot{\phi}}^2}{2H}$, $y = \frac{\sqrt{V}}{\sqrt{3} H}$ and $\lambda = -\frac{1}{{\dot{\phi}}V}\frac{dV}{d\phi}$.
\ec
For the present non-canonical scalar field model having a constant equation of state parameter $\omega$, the choice $\frac{dV}{d\phi} = -\lambda \dot{\phi} V$ leads to a simple power law potential $\left[V(\phi) = V_0 (\phi - \phi_{0})^{-4}\right]$. So the present analysis is valid for power-law potentials in particular. However other choices of potentials are also possible depending upon the dynamics of the $\phi$-field.\\
In terms of the new variables, the evolution equations for the scalar field can be written as a plane-autonomous system:
\be\label{xprime}
x^{\prime} = -Wx + \frac{3}{2}x(1 + \frac{x^2}{3} - y^2) + \lambda y^2
\ee
\be\label{yprime}
y^{\prime} = \frac{3}{2}y(1 + \frac{x^2}{3} - y^2) - \lambda xy
\ee 
where $W = \frac{2}{3}(3 - \alpha)$ and a prime indicates differentiation with respect to $N$ ($N = ln a$). \\
Also the constraint equations come out as 
\be\label{Omegaphisa}
\Omega_{\phi} = \frac{\rho_{\phi}}{3H^2} = x^2 + y^2
\ee
and 
\be
\Omega_{m} = \frac{\rho_{m}}{3H^2} = 1 - x^2 - y^2
\ee
The total EoS parameter can be written as
\be
\omega_{tot} = \frac{p_{\phi}}{\rho_{m} + \rho_{\phi}} = \frac{x^2}{3} - y^2
\ee
and the condition for acceleration is $\omega_{tot} < -\frac{1}{3}$. \\
At critical points, say $(x^*,y^*)$, both $x^{\prime}$ and $y^{\prime}$ becomes zero. In order to study the stability of the critical points, the system is perturbed about the fixed points by small amounts $u$ and $v$ as 
\be\label{xy}
x = x^* + u, ~~~y = y^* + v.
\ee
Putting these in $x^{\prime}$ and $y^{\prime}$, one obtains first order differential equations of the form  
\be
\left [
    \begin{array}{cc}
      u^{\prime} \vspace{0.07in}\\
      v^{\prime}
    \end{array}
  \right]= \cal M\left [ 
    \begin{array}{cc}
      u \vspace{0.07in}\\
      v
    \end{array}
  \right]
\ee
where 
\bc
$
\cal M = \left [
    \begin{array}{cc}
      \frac{\partial x^{\prime}}{\partial x} \hspace{0.3in} \frac{\partial x^{\prime}}{\partial y} \vspace{0.07in}\\
      \frac{\partial y^{\prime}}{\partial x} \hspace{0.3in} \frac{\partial y^{\prime}}{\partial y} \vspace{0.07in}
    \end{array}
  \right ]. 
$
\ec
$\cal M$ is a called Jacobian matrix at the fixed points. The physical stability of an autonomous system is determined completely by the eigenvalues of the matrix $\cal M$. \\
(i) If the eigenvalues are real and have opposite signs, then the fixed point is a saddle point. \\
(ii) If both the eigenvalues are real and negative, then the fixed point is stable. \\
(iii) If the eigen values are real and positive, the fixed point is unstable. \\
(iv) If the eigen values are complex but real parts of the eigen values are negative, then the fixed point is a stable spiral. \\
(v) For complex eigen values if the real parts of the eigen values are positive, then the fixed point is a unstable spiral.
A detailed analysis of the stability criteria is given in the Refs. \cite{EJC, holden, ross}.
In order to obtain a stable solution we require that all the eigenvalues of $\cal M$ must have negative real part. For the present system, the Jacobian matrix is\\ 
\bc
$
\cal M = \left [
    \begin{array}{cc}
      \left(\frac{3}{2} - W + \frac{3}{2}({x^*}^2 - {y^*}^2)\right) \hspace{0.1in} \left(2\lambda y^* - 3 {x^*} {y^*} \right) \vspace{0.07in}\\
      \left({x^*} {y^*} - \lambda {y^*}\right) \hspace{0.1in} \left(\frac{3}{2} - \lambda {x^*} + \frac{1}{2}({x^*}^2 - 9{y^*}^2)\right) \vspace{0.07in}
    \end{array}
  \right ]
$
\ec
where $W = \frac{2}{3}(3-\alpha)$.\\
The critical points and the corresponding eigen values for the present autonomous system are listed in Table 1. 
\bc
\begin{table}[!h]
\centering
\begin{tabular}{|l|c|c|c|}
\hline
&$x^*$ &$y^*$ &Eigenvalues\\
\hline
i& 0 & 0 &$\frac{3}{2} - W,~~\frac{3}{2}$ \\
\hline
ii& $\sqrt{2W - 3}$& 0 &$2W - 3$, $W - \lambda \sqrt{2W - 3}$\\
\hline
iii& $-\sqrt{2W - 3}$ & 0 &$2W - 3$. $W + \lambda \sqrt{2W - 3}$\\
\hline
iv& $b_{+}$ &$\frac{p}{2\sqrt{2\lambda}}$ & $\frac{1}{2}[(k_1 + k_2) \pm \sqrt{(k_1 + k_2)^2 - 4 k_3}]$\\ 
\hline
v& $b_{+}$ &$-\frac{p}{2\sqrt{2\lambda}}$ &$\frac{1}{2}[(k_1 + k_2) \pm \sqrt{(k_1 + k_2)^2 - 4 k_3}]$\\
\hline
vi& $b_{-}$ &$\frac{q_1}{\sqrt{2\lambda}}$  &$\frac{1}{2}[(k_4 + k_5) \pm \sqrt{(k_4 + k_5)^2 - 4 k_6}]$\\
\hline
vii& $b_{-}$ &$-\frac{q_1}{\sqrt{2\lambda}}$  & $\frac{1}{2}[(k_4 + k_5) \pm \sqrt{(k_4 + k_5)^2 - 4 k_6}]$\\ 
\hline
\end{tabular}
\caption{Critical points and the corresponding eigen values for the present system}
\label{table1}
\end{table}
\ec
Here for simplicity we have introduced the following parameters :\\ \\
$b_{\pm} = \beta \pm \frac{\sqrt{-48\lambda^2 + (3W+2\lambda^2)^2}}{8\lambda}$, $\beta=\frac{(3W+2\lambda^2)}{8\lambda}$,\\ \\
$c = \frac{3W^2}{\lambda} + 24\lambda - 4W\lambda - 4\lambda^3$,\\ \\
$p = \frac{1}{2}\sqrt{c + (b_{+} - \beta)(8W - 16\lambda^2)}$,\\ \\
$q_{1} = \frac{1}{4}\sqrt{c + ( b_{-} - \beta)(8W - 16\lambda^2)}$,\\ \\
$k_{1} = \frac{3}{2} - W + \frac{3}{2}b^2_{+} - \frac{3p^2}{16\lambda}$,\\ \\
$k_{2} = \frac{3}{2} - \lambda b_{+} + \frac{b^2_{+}}{2} - \frac{9p^2}{16\lambda}$,\\ \\
$k_{3} = k_{1}k_{2} - \frac{p^2}{8\lambda}(b_{+} - \lambda)(2\lambda - 3b_{+})$,\\ \\
$k_{4} = \frac{3}{2} - W + \frac{3}{2}b^2_{-} - \frac{3q^2_{1}}{4\lambda}$,\\ \\
$k_{5} = \frac{3}{2} - \lambda b_{-} + \frac{b^2_{-}}{2} - \frac{9q^2_{1}}{4\lambda}$,\\ \\
$k_{6} = k_{4}k_{5} - \frac{q^2_{1}}{2\lambda}(b_{-} - \lambda)(2\lambda - 3b_{-})$,\\ \\

It is not very straight forward to analyse the nature of the eigen values (positive or negative) from the above table as too many parameters are involved.
Infact the stability criteria will crucially depend on the values of $\lambda$ and $\alpha$ - the strength of the coupling factor between the DE and the DM 
components which is again dependent on the values of the model parameters $\epsilon$ and $\omega_{\phi}$.  So for simplicity, we choose the values of the
above mentioned parameters of the model as $\lambda = 1$, $\epsilon = 1.1$ and $\omega_{\phi} = -0.9$, which are the values for which we have plotted the
graphs in the previous sections. With these values chosen, we try to analyse the nature of stability of the critical points. The critical points, the nature of stability and acceleration ($\omega_{tot} < -\frac{1}{3}$) are summarized in Table 2 for the chosen values of the model parameters. 
\bc
\small{
\begin{table}[!h]
\centering
\begin{tabular}{|l|c|c|c|c|c|c|}
\hline
&$x^*$ &$y^*$ &Nature of eigenvalues  & Stability? & ${\omega}^*_{tot}$&Acceleration?\\
\hline
i& $0$ & $0$ & real, unequal and opposite signs  & Saddle point  & $0$ & No\\
& & & $(\mu_1 = -5.83, \mu_2 = +1.5)$ & & & \\
\hline
ii& $\sqrt{2W - 3}$ & 0 & real, unequal and positive  & Unstable node  & 3.88  & No\\ 
& & & $(\mu_1 = +11.66, \mu_2 = +3.91)$ & & & \\
\hline
iii& $-\sqrt{2W - 3}$ & 0 & real, unequal and positive  & Unstable node & 3.88 & No\\ 
& & & $(\mu_1 = +11.66, \mu_2 = +10.74)$ & & & \\ 
\hline
iv& $b_{+}$ & $\frac{p}{2\sqrt{2\lambda}}$ & real, unequal and opposite signs  & Saddle point & 2.87  & No\\ 
& & & $(\mu_1 = +17.98, \mu_2 = -10.91)$ & & & \\ 
\hline
v& $b_{+}$ & $-\frac{p}{2\sqrt{2\lambda}}$ & real, unequal and opposite signs & Saddle point & 2.87 & No\\
& & & $(\mu_1 = +17.98, \mu_2 = -10.91)$ & & & \\ 
\hline
vi&$b_{-}$ & $\frac{q_{1}}{\sqrt{2\lambda}}$ & real, unequal and negative & Stable node  & -0.898 & Yes\\
& & & $(\mu_1 = -3.03, \mu_2 = -6.85)$ & & & \\ 
\hline
vii&$b_{-}$& $-\frac{q_{1}}{\sqrt{2\lambda}}$ & real, unequal and negative & Stable node  & -0.898 & Yes\\ 
& & & $(\mu_1 = -3.03, \mu_2 = -6.85)$ & & & \\ 
\hline
\end{tabular}
\caption{The properties of the critical points. This is for $\lambda = 1$, $\epsilon = 1.1$ and $\omega_{\phi} = -0.9$. Here, $\alpha = 3 - \frac{\epsilon}{1 + \omega_{\phi}} = -8.0$ and $W = \frac{2}{3}(3 - \alpha) = 7.33$.}
\label{table2}
\end{table}
}
\ec
It is evident from Table 2 that only for the critical points (vi) and (vii), the system will generate a stable solution. Also it is found that an accelerating solution is possible only for these two critical points and thus our interest lies in these two points.\\
For points (vi) and (vii), the chosen values of the model parameters give\\
 $x^* = b_{-} = 0.12$ and $y^* = \pm \frac{q_1}{\sqrt{2\lambda}} \sim \pm 0.95$.\\
With these values of $x^*$ and $y^*$, equation (\ref{Omegaphisa}) readily gives\\ 
$\Omega_{\phi} \sim 0.9169$ and $\Omega_{m} \sim 0.0831$ and the deceleration parameter \\
$q = -1 - \frac{\dot{H}}{H^2} = -1 + \frac{3}{2}\left[1 + \frac{x^2}{3} - y^2\right] = -0.84.$ \\ 
\par
\begin{figure}[!h]
\centerline{\psfig{figure=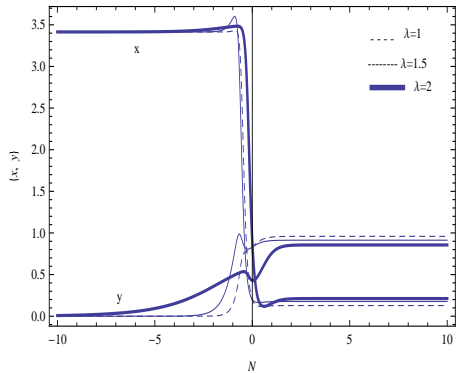,height=50mm,width=60mm}}
\caption{\normalsize{\em Plot of $x$ and $y$ against $N$ for $\lambda = 1.0, 1.5, 2.0$, $\epsilon = 1.1$ and $\omega_{\phi} = -0.9$.}}
\label{figxy}
\end{figure}

\begin{figure}[!h]
\centerline{\psfig{figure=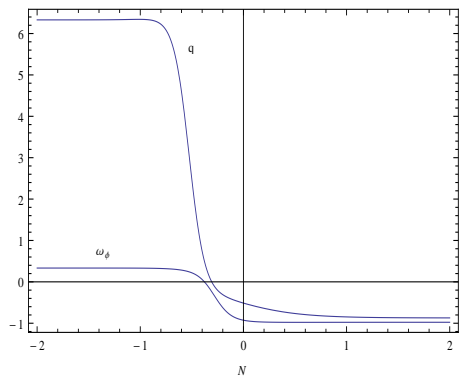,height=50mm,width=60mm}}
\caption{\normalsize{\em Evolution of parameters $q$ and $\omega_{\phi}$ for $\lambda = 1$.}}
\label{qsa}
\end{figure}

\begin{figure}[!h]
\centerline{\psfig{figure=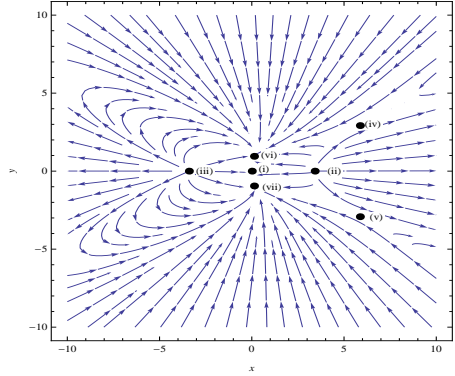,height=50mm,width=60mm}}
\caption{\normalsize{\em Phase diagram of the autonomous system in the x-y plane for $\lambda = 1$, $\epsilon = 1.1$ and $\omega_{\phi} = -0.9$.}}
\label{figps}
\end{figure}

\begin{figure}[!h]
\centerline{\psfig{figure=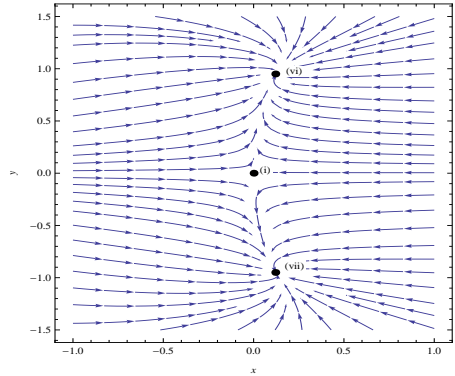,height=50mm,width=60mm}}
\caption{\normalsize{\em Phase diagram of the autonomous system in the x-y plane near the stable points.}}
\label{figzoom}
\end{figure}

 These values are little bit higher than the ones suggested by recent observations, but it has been found that as we increase the value of $\lambda$, the calculated values for various parameters of the model approach the observationally suggested values. It deserves mention that these values provide us information about various cosmological parameters at the stable fixed points (vi) and (vii) for the chosen values of $\lambda$, $\epsilon$ and $\omega_{\phi}$ and does not provide us with the complete evolution of the universe. However the evolution of the scalar field (DE) can be obtained by numerically solving equations (\ref{xprime}) and (\ref{yprime}). The behaviour of $x$ and $y$ against $N$ for different values of $\lambda$ are shown in figure \ref{figxy}. It is evident from figure \ref{figxy} that the evolution of the system is not very sensitive to the value of $\lambda$ and the at the early phase of evolution ($N<0$) the kinetic term of the scalar field (demonstrated by parameter $x$) was dominating and at present the potential term is dominating over the kinetic term, as expected for a scalar field model of dark energy with $\omega_{\phi} > -1$. This reassures the results obtained in the previous section. An almost similar result has been obtained for slight variation in the values of $\epsilon$ and $\omega_{\phi}$ also.\\  
The evolution of the deceleration parameter $q$ and the equation of state parameter $\omega_{\phi}$ against $N$ for $\lambda =1$ is shown in figure \ref{qsa}.  It is evident from figure \ref{qsa} that the universe was undergoing a decelerated expansion phase and enters into an accelerated phase in the recent past and the evolution of $\omega_{\phi}$ indicates that the value of $\omega_{\phi}$ was positive initially, at present it is close to $-0.9$ and settles to a value $-1$ in future. Also it is seen that the variation of $\omega_{\phi}$ over the total span is very small and thus it is justified to consider $\omega_{\phi}$ constant over the entire span of evolution. Furthermore as $\omega_{\phi}$ settles to a value close to $-1$ and never crosses $-1$, this indicates that the present model will behave like a $\Lambda$CDM model in future and the universe will evolve to the asymptotic de Sitter space-time. This is in agreement with the various observational results \cite{wobs1, wobs2, wobs3}. As  $\omega_{\phi}$ or $\omega_{tot}$ never crosses $-1$, this indicates that the non-canonical scalar field considered in the present model is not phantom and thus there is no future singularity in this model \cite{phantom, phantom1}.\\
One can also draw the phase portrait for the system for the chosen values of the model parameters as shown in figure \ref{figps} which shows that the 
points (ii) and (iii) are unstable fixed points and (vi) and (vii) are stable fixed points. In figure \ref{figzoom} we zoom the plot around the points (vi) and (vii) to understand the stability a bit more clearly. So the universe can start its evolution from any of the unstable points and will finally settle down into one of the stable configurations (vi) or (vii).   
\section{Conclusion}
In this paper, we have described a cosmological model with a non-canonical scalar field in which the scalar field is allowed to interact with the matter component of the universe. We have considered a specific form for the coupling function as $Q = \alpha H {\dot{\phi}}^{4}$. As mentioned earlier, the form of $Q$ chosen is quite arbitrary, but as nothing specific is known about the nature of {\em dark energy}, any coupling term can be considered phenomenologically. Next we have obtained analytical solutions for various cosmological parameters of the model with a constant equation of state parameter for the $\phi$-field. This toy model is somewhat restricted in this sense but observational data suggests that the allowed values of the equation of state parameter is $-1.1 \le \omega_{\phi} \le -0.9$ \cite{vasey, davis}. Considering this small range of allowed values, it is quite justified to consider $\omega_{\phi}$ as a constant parameter. However, it would be interesting to study the properties of the model with a varying equation of state parameter for the scalar field.\\ 
Furthermore, it has been found that for this interacting model, the deceleration parameter $q$ undergoes a smooth transition from a decelerated to an accelerated phase of expansion driven by the non-canonical scalar field $\phi$ (see Figure \ref{figq1}). This is essential for the structure formation of the universe.\\ 
In the coupling term $Q$, the parameter $\alpha$ was initially kept arbitrary and its value has been found out from other parameters of the model. It has been found that the $\alpha < 0 $ which eventually shows that the flow of energy is from the dark energy sector to the dark matter sector but still the dynamics of the universe eventually makes the energy density for the DE sector to grow at late times to drive the acceleration of the universe. Therefore, this interaction scenario could lead to the solution of the coincidence problem.\\
We have also investigated the stability of the interacting non-canonical scalar field model. It has been found that for a particular choice of model parameters (consistent with the analytical model described earlier), there are only two physically relevant generic stable fixed points, which can provide an accelerating solution. The stability conditions, however, depend totally on the choice of model parameters and may be different for different choices of parameters. However, for the present choice of model parameters it has been found that the universe will settle down to a $\Lambda$CDM model in future and there will not be any future singularity in the non-canonical interacting dark energy model. However here the analysis has been done for a particular choice of interaction and a wide range of possibilities are open for various choices of interaction terms.  
\section{Acknowledgment}
One of the authors (AAM) acknowledges UGC, Govt. of India for financial support through Maulana Azad National Fellowship. SD wishes to thank IUCAA, Pune for the associateship programme where part of this work has been carried out.  
                                         
\end{document}